\documentclass[prl, aps, twocolumn,showpacs]{revtex4-1}
\usepackage{graphicx}
\usepackage{dcolumn}
\usepackage{bm}
\usepackage{color}
\usepackage{amssymb,amsmath}

\begin{document}
\title{Deconvoluting chain heterogeneities from driven translocation through a nano-pore}
\author{Ramesh Adhikari}
\author{Aniket Bhattacharya}
\altaffiliation[]
{Author to whom the correspondence should be addressed}
\email{aniket@physics.ucf.edu}
\affiliation{Department of Physics, University of Central Florida, Orlando, Florida 32816-2385, USA}
\date{\today}

\begin{abstract}
We study translocation dynamics of a driven compressible semi-flexible chain consisting of alternate
blocks of stiff ($S$) and flexible ($F$) segments of size $m$ and $n$ respectively for different
chain length $N$ in two dimension (2D).
The free parameters in the model are the bending rigidity $\kappa_b$ which controls the three body
interaction term, the elastic constant $k_F$ in the FENE (bond) potential between successive monomers,
as well as the segmental lengths $m$ and $n$ and the repeat unit $p$
($N=m_pn_p$) and the solvent viscosity $\gamma$. We demonstrate that due to the change in entropic barrier and
the inhomogeneous viscous drag on the chain backbone a variety of scenarios are possible amply manifested in the waiting
time distribution of the translocating chain. These information
can be deconvoluted to extract the mechanical properties of the chain at various length scales and
thus can be used to nanopore based methods to probe bio-molecules, such as DNA, RNA and proteins.
\end{abstract} 
\maketitle
Polymer translocation (PT) through a nano-pore (NP) is being explored for more than a decade as
a NP based device has the potential to provide single molecule detection when a DNA is driven
electrophoretically through a NP~\cite{Muthukumar2011,Milchev2011}. Unlike traditional Sanger's method~\cite{Sangers}
this does not require amplification; thus
one can in principle analyze a single genome~\cite{Mardis_Nature_2011}. Progress towards this target offers challenges to overcome which
have attracted a lot of attention from various disciplines of sciences and engineering~\cite{Movileanu_SM_2008,Bayley_NN_2013}.
A large fraction of  theoretical and numerical studies have been devoted to
translocation studies of flexible homo-polymers~\cite{Muthukumar2011,Milchev2011}. However, to extract sequence specific
information for a DNA or a protein, as they translocate and/or unfold
through a nanopore, one needs generalization of the model to account for how different segments of the translocating
polymer interact
with the pore or the solvent. Translocation of the heterogeneous polymer has been studied in the past for a fully flexible polymer
where different segments encounter different forces~\cite{Luo_JCP_2007,Luo_PRL_2008,Gauthier_JCP_2008,Mirigan_JCP_2012}.
For periodic blocks one observes novel periodic fringes from which information about the
block length can in principle be readily extracted~\cite{Luo_JCP_2007,Luo_PRL_2008}. Recently,
de Haan and Slater~\cite{Slater_PRL2013} have studied translocation of rod-coil polymer
through a nanopore in the {\em quasi-static limit} (weakly driven through narrow pore and negligible fluid viscosity).
They have used incremental mean first passage time (IMFPT)~\cite{IMFPT} approach and verified that in the quasi-static
limit the stiff and flexible segments can be discriminated due to local entropic mismatch between the stiff and
flexible segments reflected in the
steps and plateaus of the IMFPT of different segments.
\par
In this letter we provide new insights for the driven heterogeneous PT through a NP where
heterogeneity is introduced by varying both the {\em bond bending} as well as the {\em bond stretching} potentials.
We study the translocation dynamics in the presence of large fluid viscosity and strong driving force so that the system
is not in the quasi-static limit as in Ref.~\cite{Slater_PRL2013}.
Our studies are motivated
by the observation that many bio-polymers, such as DNA and proteins exhibit helical and random coil segments whose elastic and
bending properties are very different, so is the entropic contribution due to very different number and nature of polymeric
conformations. It is also likely that a double stranded ($ds$) DNA can be in a partially melted state whose coarse-grained description
will require nonuniform bond bending and bond-stretching potentials for different regions.
As a result, if one wants to develop a NP based device to detect and identify the translocating
segments, a prior knowledge of their residence inside the pore will be extremely useful.
Naturally, the length scale of the heterogeneity $\xi(n,m)$, where $m$ and $n$ are the lengths
of the stiff and flexible segments respectively in each block, will obviously be an important parameter for the analysis of the translocation
problem. Thus, we first show that
a proper coarse graining  of the model in units of $\xi$ will lead to the known results for the
homopolymer translocation. Then we further analyze the results at the length scale of the blob size $\xi$ and show how
the chain elasticity and the chain stiffness introduce fine prints in the translocation process.
We explain our findings using Sakaue's non-equilibrium tension propagation (TP) theory~\cite{Sakaue_PRE_2007} recently verified
for a CG models of semi-flexible chain by us~\cite{Bhattacharya_JPolyC_2013,Adhikari_JCP_2013,comment1}.
We have used Lennard-Jones (LJ), Finitely Extensible Nonlinear Elastic (FENE) spring potential and
a three body bond bending potential to mimic excluded
volume (EV), bond stretching between two successive monomers, and stiffness of the chain respectively,
and applied a constant external force ($F_{ext} = 5.0$) at the pore in the translocation direction.
We have used the Brownian dynamics (BD)
scheme to study the heterogeneous PT problem. The details of the BD methods are the same as
in our recent publications~\cite{Bhattacharya_JPolyC_2013,Adhikari_JCP_2013}. Initially we keep the elastic spring constant ($k_F$)
to be the same throughout the chain and choose the bending stiffness $\kappa_b  = 0$ and 16.0 for the fully flexible and the
stiff segments respectively.
Later we show that by making the elastic potential for the relatively more flexible part weaker one can reverse the relative
friction on the chain segments which results in novel waiting time distributions serving as the fingerprint of the structural motifs
translocating through the pore. \par
\begin{figure}[ht!]              
\begin{center}
\includegraphics[width=0.7\columnwidth]{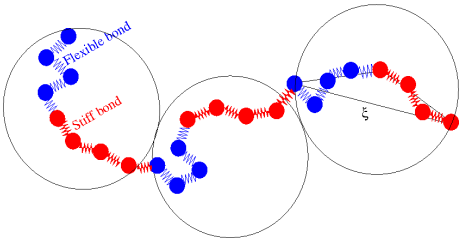}
\end{center}
\vskip -0.5truecm
\caption{\small Blob model of a polymer chain of chain length (N = 24) and segmental length (m = 4).
Each repeated unit can be considered as a single blob of length $\xi \sim m^{\beta}$. Please see the text below.
}
\label{model}
\end{figure}
\begin{figure}[ht!]                
\begin{center}
\includegraphics[width=0.7\columnwidth]{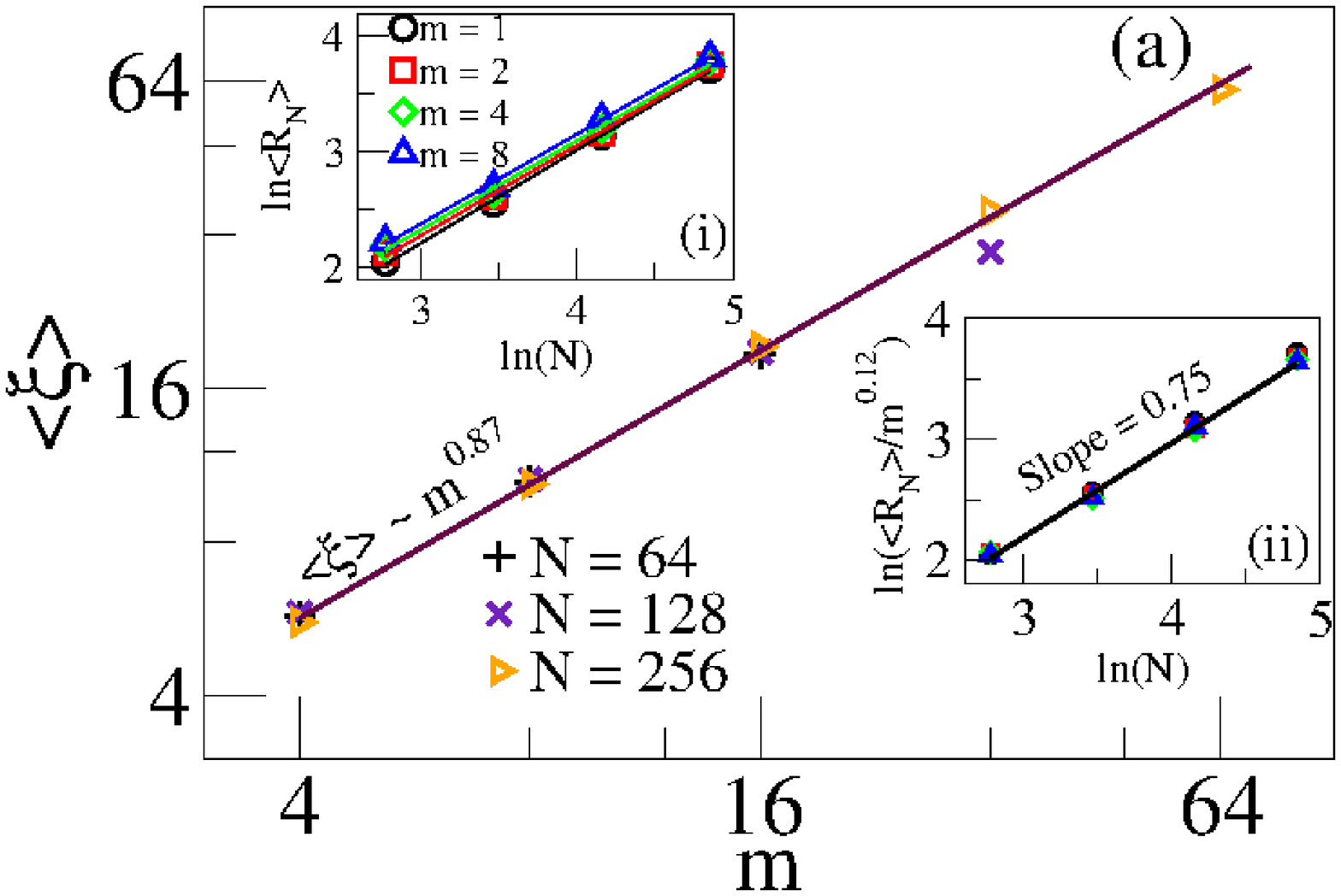}
\includegraphics[width=0.7\columnwidth]{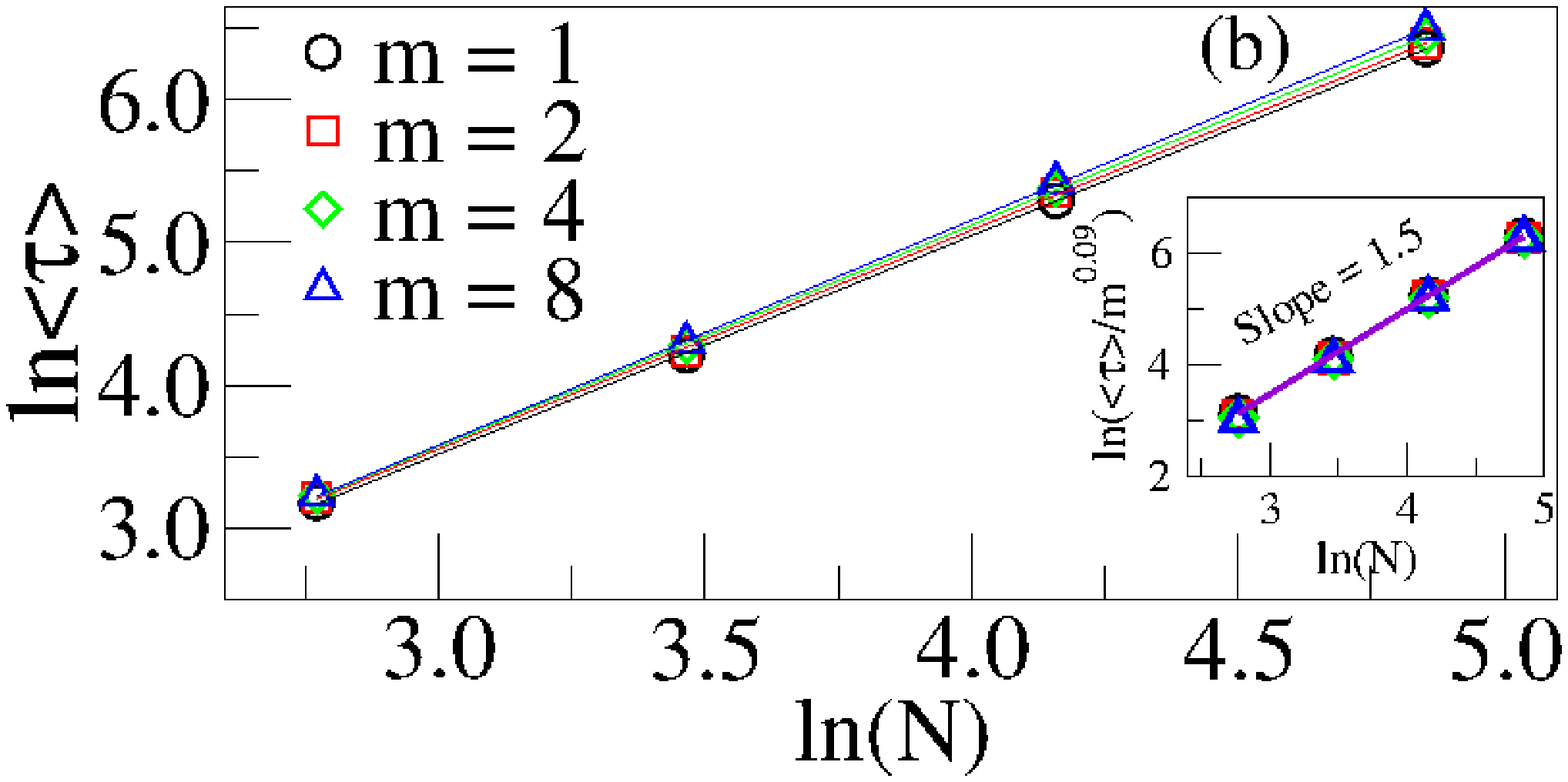}
\end{center}
\vskip -0.5truecm
\caption{\small (a) Log-log plot of blob-size $\langle\xi\rangle$ as a function of $m$ for
N = 64 (black plus), N = 128 (violet cross) and N = 256 (orange right-triangle). The solid line represents $\langle\xi\rangle \sim  m^{0.87}$. Insets: (i) log-log plot of $\langle R_N \rangle$
as a function of $N$ for different $m$,
(ii) collapse of $\langle R_N \rangle /m^{0.12} \sim N^\nu$ on the same master plot.
(b) Log-log plot of $\langle \tau \rangle$ as a function of $N$ for different $m$.
Inset: scaling and collapse of $\langle \tau \rangle /m^{0.09} \sim N^{2\nu}$.}
\label{blob}
\end{figure}

$\bullet$~{\em Blobsize and scaling}:~We consider heterogeneous chains consisting of alternate symmetric ($m=n$) periodic blocks of stiff and flexible
segments of $m$ monomers so that the block-length is $2m$ ($m$ = 1,2,3,4) as shown in Fig.~\ref{model}.
First we investigate how the alternate stiff and flexible segments
of equal length affect the end-to-end distance $\langle R_N(m)\rangle $ and the mean first passage time (MFPT) as a function of the periodic
block-length (Fig.~\ref{model}), compared to a homo-polymer of equal contour length $N$. To a first approximation one
can think of this chain as a
flexible chain of $N/2m$ segments, of certain blob size $\xi$. The blob size $\xi$ in general will be a function of the
block-length and bending rigidity of the flexible and stiff segments.
For our particular choice of bending rigidity for the flexible ($\kappa_b=0$) and stiff ($\kappa_b=16$)
segments from simulation results for $N=64-256$ we find an expected power law scaling
$\xi \sim m^{\beta}$ where $\beta = 0.87$ (Fig.~\ref{blob}). Obviously the exponent $\beta$ is
non-universal as it depends on $\kappa_b$ and $k_F$, but the universal aspects of the entire chain can be
regained through scaling with $\xi$ as shown in Fig.~\ref{blob}.
The conformation statistics of this basic unit $\xi$ controls both the conformation and
translocation properties of the entire chain as follows. We can write
$ \langle R_N \rangle \equiv \langle \sqrt{R_N^2} \rangle \sim  \langle \xi \rangle (N/2m)^{\nu} \sim m^{\beta}N^{\nu}/m^\nu$, where
$\nu$ is the Flory exponent. This implies
$ \langle R_N \rangle/N^\nu \sim m^{\beta-\nu} = m^{0.12}$ (where $\nu=0.75$ is the Flory exponent in 2D). Simulation data in the insets of Fig.~\ref{blob}(a) confirms our scaling prediction. Likewise,
we show that the MFPT $\langle\tau\rangle/N^{2\nu} \sim m^{0.09}$. For small $N$ it has been found earlier that
$\langle \tau \rangle \sim \langle R_N \rangle /N^{-\nu} \sim N^{2\nu}$~\cite{Luo_JCP_2nu}. Therefore, as expected by proper coarse graining by the
elemental block we get back the results for the fully flexible chain. We now show how characteristics of translocation
are affected by the chain heterogeneity. \par
\begin{figure}[ht!]          
\begin{center}
\includegraphics[width=0.7\columnwidth]{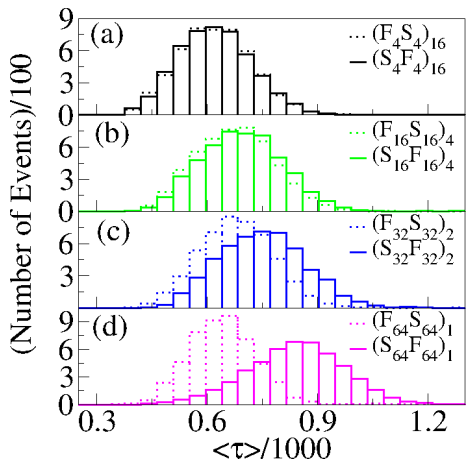}
\end{center}
\vskip -0.5truecm
\caption{\small Histograms of first passage time for chain length $N = 128$ and segmental length (a) $m$ = 4,
(b) $m$ = 16, (c) $m$ = 32, and (d) $m$ = 64. The dotted/solid lines represent the flexible($F_mS_m$)/stiff($S_mF_m$) segment entering
the pore first. For larger block size the effect of order of entry is clearly visible.}
\label{hist}
\end{figure}

\begin{figure}[ht!]          
\begin{center}
\includegraphics[width=0.7\columnwidth]{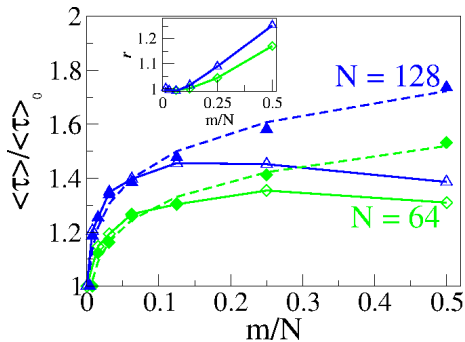}
\end{center}
\vskip -0.5truecm
\caption{\small MFPT (scaled by the MFPT of respective flexible homo-polymer) for chains $F_mS_m$ and $S_mF_m$ as a function of $m/N$
for  chains $N=64$ (green diamonds) and $N=128$ (blue up-triangles). The open/closed symbols correspond to
flexible/stiff segment entering the pore first. The inset shows the ratio of the MFPT for
$SF$ to $FS$ orientation. The nanopore is capable of differentiating if a flexible (F) block or a stiff (S)
block entered the pore first.}
\label{tau}
\end{figure}

$\bullet$~{\em Effect of chain heterogeneity on translocation}:~In presenting the results
we use the notation $(F_mS_n)_p$/$(S_mF_n)_p$ to denote $p$ blocks of an ordered flexible/stiff and stiff/flexible segments of length $m$ and
$n$ respectively ($N=(m+n)p$)
and that the flexible/stiff segment enters the pore first. Fig.~\ref{hist} and Fig.~\ref{tau}
reveal quite a few novel results that we explain using TP theory. For small block-length the order in which the chain enters the
pore (either stiff or flexible segment) neither make a big difference in the shape of the histogram  (Fig.~\ref{hist}(a))
nor in the MFPT (Fig.~\ref{tau}). For larger block lengths the difference
between the histograms for $S_mF_m$ and $F_mS_m$ are quite clear and the dependence of $\tau$ on $m$ are also different
as seen in Fig.~\ref{tau}.  For the case when the stiff portion enters
the pore first, the MFPT monotonically increases but in the other case it shows a maximum (Fig.~\ref{tau}).
We now explain this in terms of our recent analysis of the translocation of semi-flexible
chain using TP theory where we showed that a stiffer chain takes a longer time to
translocate~\cite{Bhattacharya_JPolyC_2013,Adhikari_JCP_2013,comment1}.
When the block lengths are small, TP gets intermittently
hindered as the tension propagates through alternate stiff and flexible regions. For longer blocks tension can
propagate more effectively unhindered for a longer time.
Therefore, when a long stiff segment enters the pore first it increases the MFPT.
But, when a long flexible segment enters the pore first it decreases the MFPT. This results a maximum
in the $\langle\tau\rangle/\langle\tau\rangle_0$ {\em vs.} $m$ curve for the FS orientation.
The difference of MFPT for $S_mF_m$ and $F_mS_m$ becomes maximum when $m=N/2$.
For relatively longer block lengths it makes a big difference in MFPT.\par
\begin{figure}[ht!]       
\begin{center}
\includegraphics[width=0.7\columnwidth]{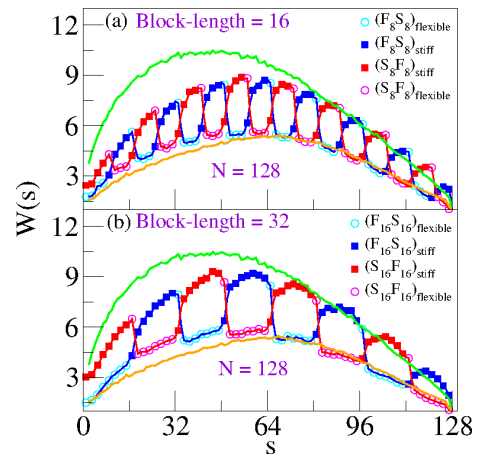}
\end{center}
\vskip -0.5truecm
\caption{\small Waiting time distribution for a $N = 128$ chain with the block-length (a) 16 and (b) 32. Azure (open circles) and
Blue (filled squares) correspond to the flexible and stiff segments when the {\em flexible segment enters the pore first ($F_mS_m$)}.
Magenta (open circles) and Red (filled squares) correspond to the flexible and stiff segments when {\em the stiff segment enters
the pore first ($S_mF_m$)}. The solid green and orange lines correspond to the waiting
time distributions for the corresponding stiff ($\kappa_b = 16.0$) and fully flexible ($\kappa_b = 0.0$) homo-polymers respectively.}
\label{res}
\end{figure}
$\bullet$~{\em Waiting time distribution}:~The total time spent by a monomer inside the pore is defined as
its waiting time $W(s)$, where $s$ is the index of monomer inside the pore (translocation coordinate).
Sum of the waiting time for all monomers is the MFPT {\em i.e.}
$\sum_{s=1}^N W(s) = \langle \tau \rangle$.
The effect of TP in stiff and flexible parts becomes most visible in the waiting time
distribution of the individual monomers of the
chain as shown in Fig.~\ref{res}. We notice that the envelopes for the corresponding homo-polymers for
a fully flexible chain ($\kappa_b=0$, solid orange line) and for the stiffer chain ($\kappa_b = 16$, solid green line) respectively
serve as bounds for the heterogeneous chains~\cite{diffraction}. As explained in our previous publication
~\cite{Adhikari_JCP_2013} the TP time corresponds to the maximum of these curves and shifts toward a lower $s$ value for a stiffer chain.
Bearing this in mind we can reconcile the fringe pattern in the light of the TP theory. The pattern has the following
features: (i) The number of fringes is equal to the
number of blocks. This is because on an average stiffer portions take longer time to translocate.
(ii) The fringes for $S_mF_m$ and $F_mS_m$ are out of phase for the same reason. (iii) The chain heterogeneity affects
the waiting distribution most at an early time; beyond the largest TP time
({\em i.e.}, the peak position of the envelope for $\kappa_b=0$) the waiting time of the individual monomers (excepting which are at the border
separating the stiff and flexible segments) becomes identical to that of the corresponding homogeneous chain. This again exemplifies to analyze
the driven translocation as a {\em pre} and {\em post} TP events. Please note that the maxima of the $W(s)$ for the heterogeneous chain lie
in between the maxima for the corresponding homogeneous cases. \par
\begin{figure}[ht!]
\begin{center}
\hskip -0.9truecm
\includegraphics[width=0.63\columnwidth]{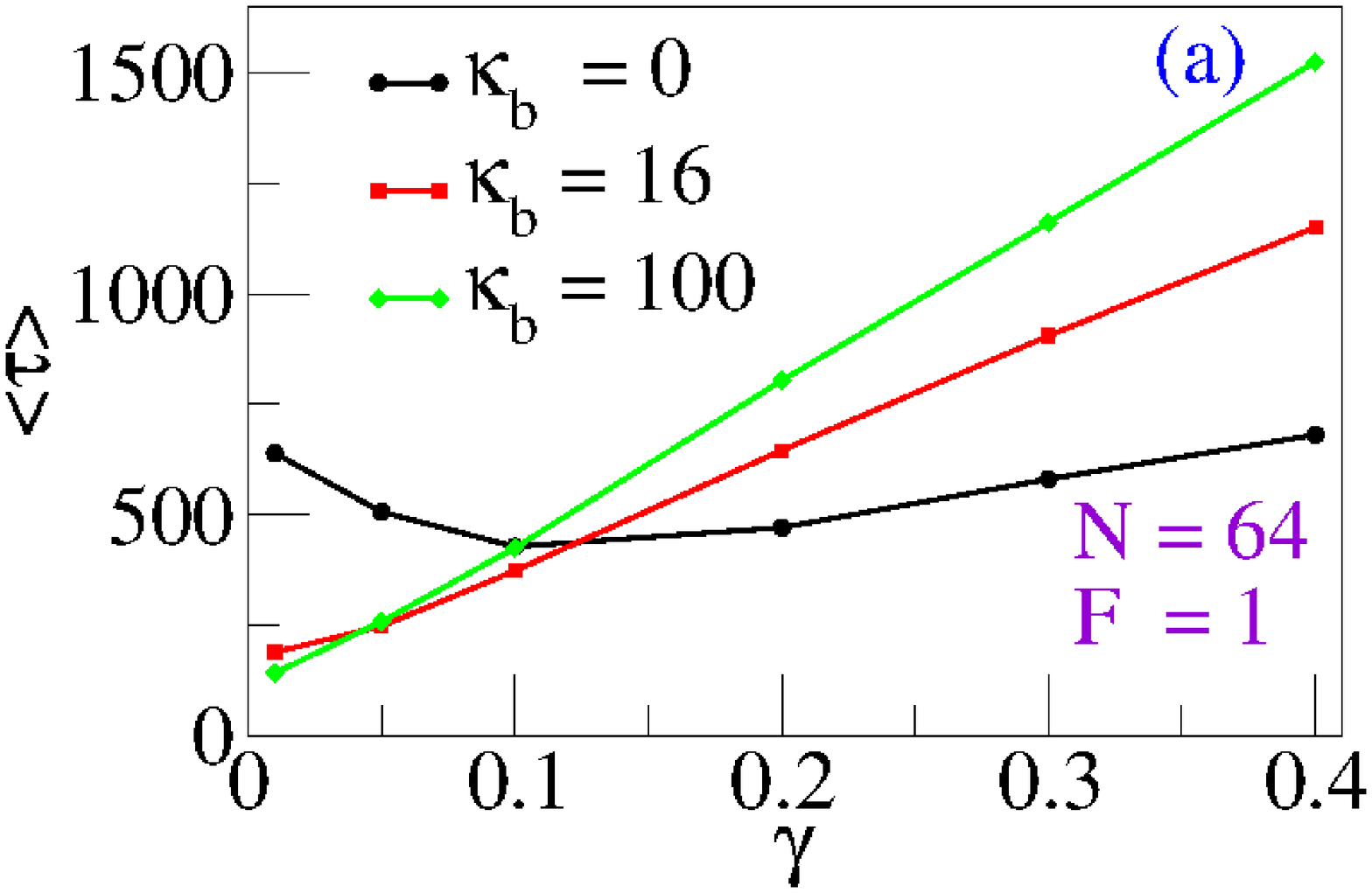}
\includegraphics[width=0.7\columnwidth]{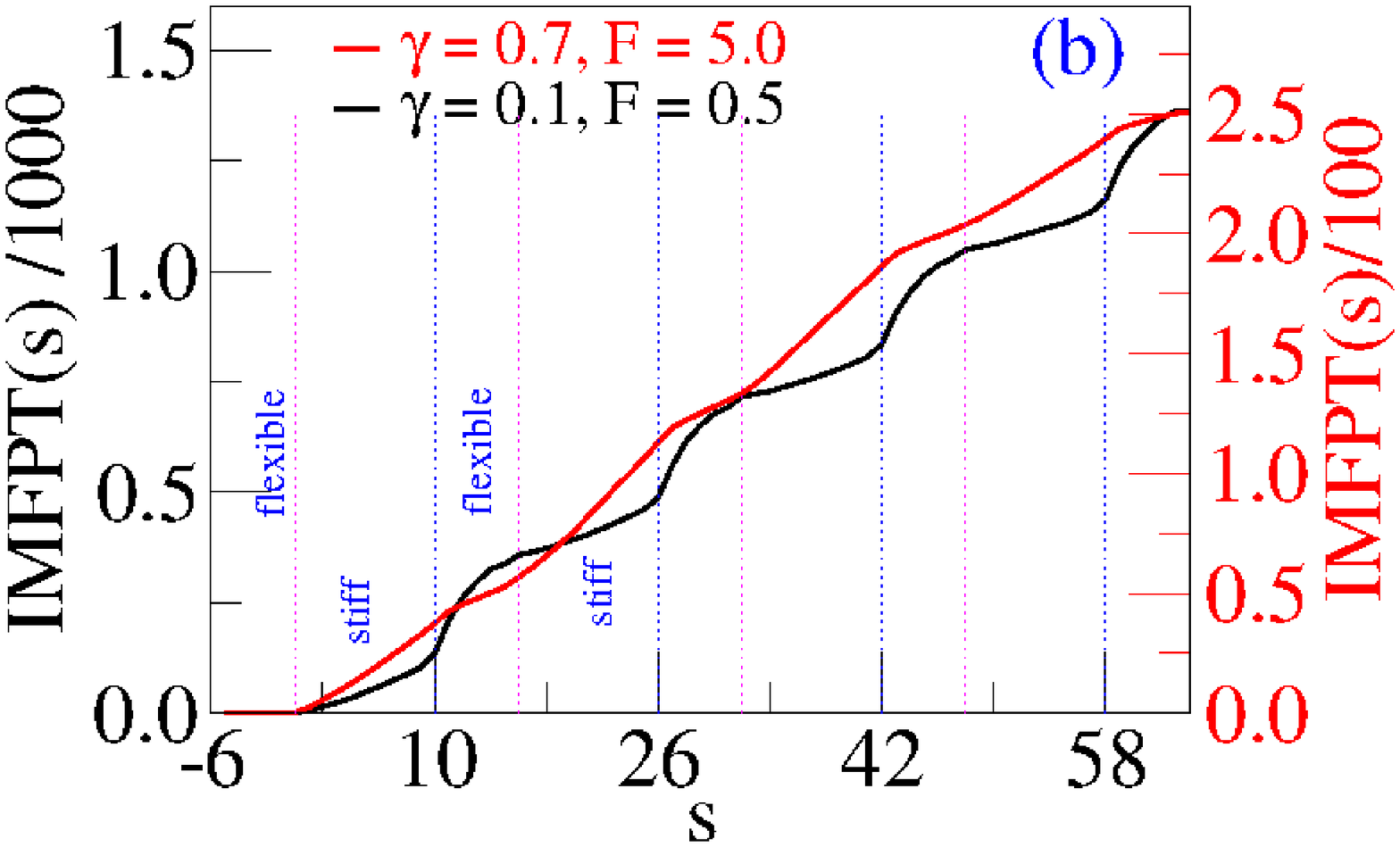}
\includegraphics[width=0.7\columnwidth]{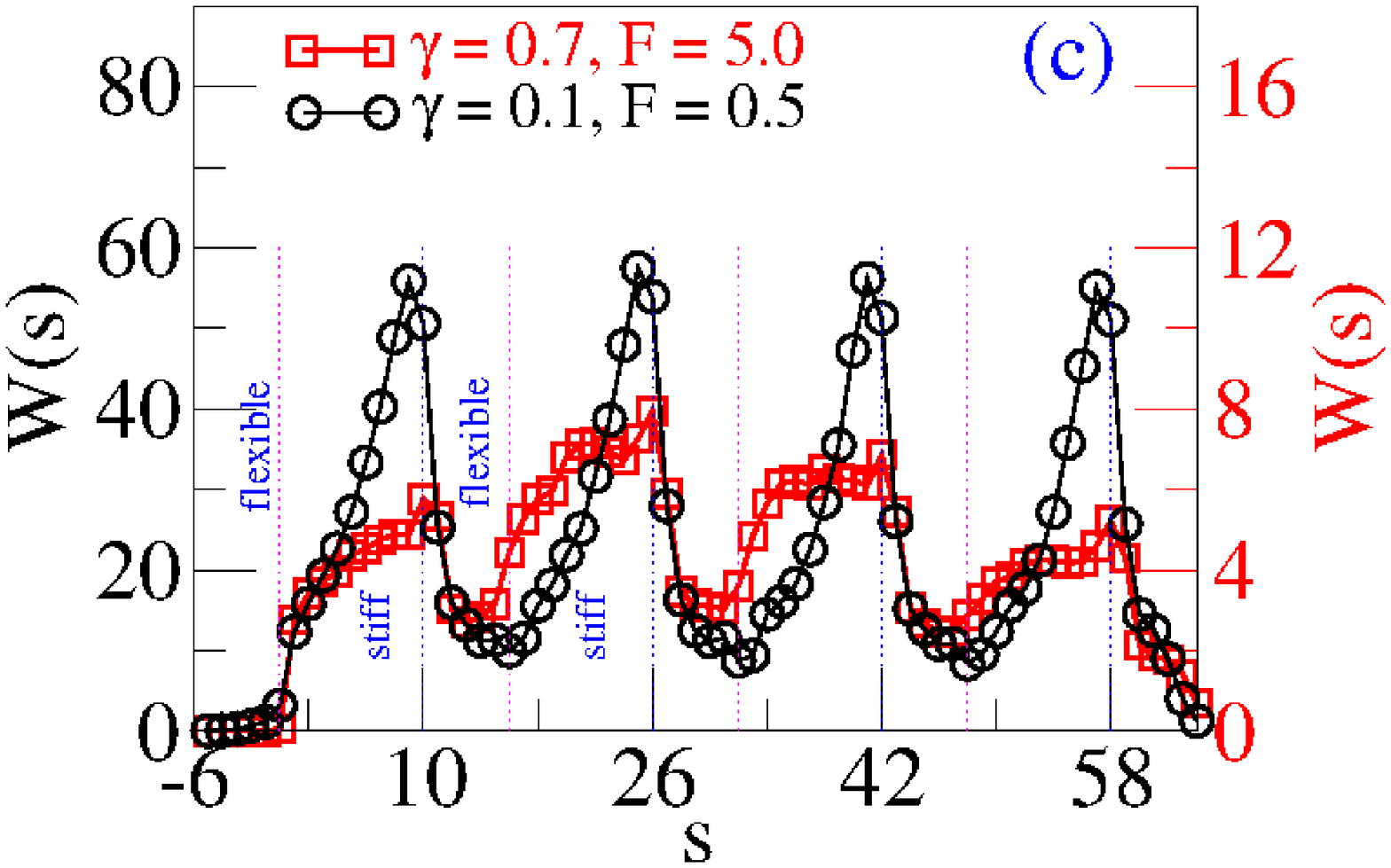}
\caption{\small (a) The MFPT for flexible and semi-flexible homopolymers of length $N = 64$ as a function of solvent-monomer friction $\gamma$.
(b) The IMFPT and (c) the waiting time distribution as a function of $s$-coordinate for a
chain ($N = 70$) in 3D with four
stiff segments ($\kappa_b = 100$) each of length ($m = 10$) and five flexible segments ($\kappa_b = 0$) each of length
($n = 6$) provided that (for (b) and (c)) the first
flexible segment is already in the $trans$-side at $t = 0$.}
\label{viscous}
\end{center}
\end{figure}
$\bullet$~{\em Effect of friction and driving force}:~In Fig.~\ref{res} we chose a value of the solvent friction associated to each monomer $\gamma = 0.7$
for which we find that a stiffer segment translocates slower through the pore. We now discuss how a variation of the solvent friction
will affect this conclusion.
We first show that the MFPT of a homopolymer of certain length
exhibits a crossover as one
varies the solvent viscosity (Fig.~\ref{viscous}(a)). It is only for extremely small $\gamma$  ({\em quasi-static limit}) the stiffer segment translocates faster as studied in ~\cite{Slater_PRL2013}.
We also have reproduced the result for a particular set of parameters (black line in Fig.~\ref{viscous}(b)).
We have shown 3D (instead of 2D) data in Fig. 6(b) and (c) only for better resolution.
This crossover effect can be explained using Sakaue's tension propagation (TP)
theory~\cite{Sakaue_PRE_2007}.
When we use a larger value of $\gamma$ (implying stiffer segment translocates slower) and/or a bias $F$
the IMFPT changes qualitatively (Fig.~\ref{viscous}(b)), which is more prominently seen
in the waiting time distribution of the individual monomers (Fig~\ref{viscous}(c)).\par
Using formulae for solvent friction from the bulk $\Gamma_{\rm solv} = \gamma N^\nu$ and pore friction $\Gamma_{\rm pore} \sim \frac{A_{pore}}{d-1}+p\gamma$
which have been discussed in Ref.~\cite{Ikonen_JCP2012,Ikonen_EPL2013}
we have checked that $\gamma = 0.7$ and  $\gamma = 0.1$ (for the chain lengths used in our simulation) correspond to solvent
dominated and pore friction dominated regimes respectively.
At high $\Gamma_{\rm solv}$, de-Haan and Slater showed that the MFPT increases linearly
with $\gamma$~\cite{deHaan_memory} for a fully flexible chain.
We see the same trend to be valid also for semi-flexible chains, albeit beyond a critical value (Fig.~\ref{viscous}(a)).
But at low $\Gamma_{\rm solv}$, the dependence of MFPT on $\gamma$ becomes non-monotonic and
it exhibits a minimum for $\gamma=\gamma_m$~\cite{Unpublished}.
This $\gamma_m$ marks the onset of change in the qualitative behavior of IMFPT or the
waiting time distribution of the individual monomers.\par
In quasi-static limit,  significantly larger local entropic barrier of a
``coil'' segment causes longer residence time.
This effect is reflected as steps in the IMFPT (Fig.~\ref{viscous}(b)) and peaks in the waiting
time distribution (Fig.~\ref{viscous}(c)).
But for the non-equilibrium situation, when the stiffer segment enters the pore, tension propagates faster along
the chain backbone~\cite{Sakaue_PRE_2007,Adhikari_JCP_2013} and more monomers in the $cis$-side set in motion. For large solvent friction this may
produce larger viscous drag dominating over local
entropic barrier resulting in the stiffer segments translocating
slower than the flexible segment. In this case the peaks in the waiting time distribution disappear (red color in Fig.~\ref{viscous}(c)).
Accordingly, one sees qualitative changes
in the corresponding IMFPT (red color in Fig.~\ref{viscous}(b)).
Therefore,  the relative fast/slow translocation of rod /coil segments through the nanopore depends on the
relative values of pore friction, solvent friction, and applied bias. \par
\par
\begin{figure}[ht!]
\begin{center}
\hskip -1.0truecm
\includegraphics[width=0.65\columnwidth]{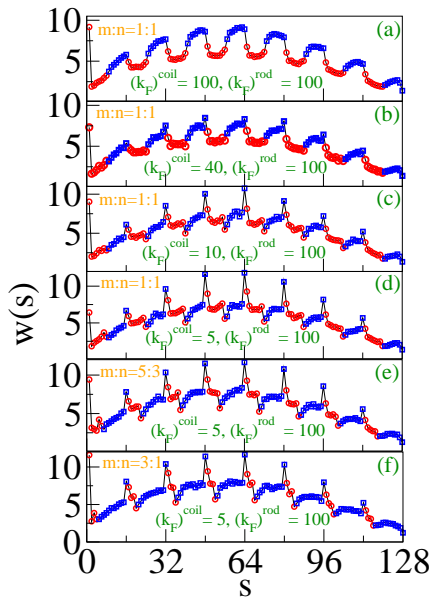}
\caption{\small The waiting time distribution as a function of $s$-coordinate for a chain ($N = 128$) with variable
$k_F$ and stiff-flexible segmental length ratio ($m/n$).
The bending stiffness $\kappa_b$ for flexible (red circles) and stiff (blue squares)
segments are 0 and 16 respectively.
The elastic stiffness ($k_F$) is 100 for stiff segments [(a)-(f)]. For flexible segments (a) $k_F = 100$
(b) $k_F = 40$ (c) $k_F = 10$ (d) $k_F = 5$ (e) $k_F = 5$ and (f) $k_F = 5$. The stiff and flexible segments are of equal length except in (e) $m\colon n = 5\colon 3$ and (f) $m\colon n = 3\colon 1$.}
\label{fene}
\end{center}
\end{figure}

$\bullet$~{\em Heterogeneous chain with a variable spring constant}:~
Finally we have extended these studies to see the consequences of allowing the elastic potential between the successive beads to
be different in each block. This situation may occur when individual building blocks are connected by linkers of different
elasticity.
Fig.~\ref{fene} shows the various combination of the spring constants $k_F$ for the heterogeneous chain.
The first four graphs Fig.~\ref{fene}(a)-(d) correspond to the waiting time distribution for the chain with
equal number of monomers in each of the flexible and stiff segments. Fig.~\ref{fene}a is the graph where
all the $F$ and $S$ segments have the same $k_F=100$ qualitatively similar to Fig.~\ref{res}. In Figs.~\ref{fene}(a)-(d)
one can see the effect of reduced value of $k_F$ for the flexible portion only.\par
Figs.~\ref{fene}(e)-(f) represent the waiting time distributions for the unequal
length of the flexible and stiff segments.
The flexible segment, being shorter, looses the conformational entropic height but the contribution
of the FENE force in the direction of translocation is enhanced. We can see the effect
of this enhancement in the increased back and forth motion (low frequency phonons of larger amplitude to softer bonds) of the chain towards the translocation
direction. The smaller is the value of $k_F$ the larger will be the amplitude
of the phonons mode which results in a longer translocation
time. Therefore, when we reduce the strength of the FENE interaction for the coil, the
coil translocates slower and we got the waiting time distribution picture
inverted for the stiff and flexible segments as seen from a comparison
of Fig.~\ref{fene}(a) to Fig.~\ref{fene}(d). This will be most prominent if
the stiff segments were chosen as rigid rods.\par
Fig.~\ref{fene}(c)-(f) show the end monomer of each semi-flexible segment has a larger waiting time. This indicates a
larger barrier height for the flexible segments. Once the barrier is overcome by the first monomer of the flexible
segment, all the following monomers of the flexible segments pass through the pore faster. The end monomer of the
flexible segment and the first monomer of the stiff segment have the lowest waiting time which means that they
have negligible barrier to overcome. Furthermore a visual comparison of Fig.~\ref{res} and Fig.~\ref{fene} shows that
the origin of the details of the waiting time distributions possibly be differentiated by a spectral decomposition analysis of the waiting time distribution.\par
To summarize, we have demonstrated how a nanopore can sense structural heterogeneity of a bio-polymer driven through a nanopore.
Not only do monomers belonging to the flexible and stiff part exhibit different waiting time distributions, we have also demonstrated
how a nanopore can sense which end of the polymer enters the pore first. Translating this information for a dsDNA will
imply that the nanopore can differentiate the 3-5 or 5-3 ends of a translocating DNA. We have explained these
results using the concepts of TP theory. We have clearly demonstrated how the fluid viscosity and an external bias can affect the relative
speed of the stiff and flexible segments. Furthermore,
unlike previously reported studies~\cite{Slater_PRL2013} we, for the first time, analyzed the interplay of the effects of
polymer heterogeneity caused by the variation of elastic and bending stiffness. We have demonstrated that
softer elastic bonds raise the MFPT~\cite{FENE1d}.
Therefore, an increase in waiting time for a stiff segment can be compensated by the waiting time for a flexible segment but having softer
elastic bonds. This observation can be exploited to tune to control the passage of polymers through NP.
It is interesting to note from Fig.~\ref{fene} that the variation in waiting time distribution arising
out of the bending stiffness variation and bond length variation can be differentiated.
Therefore, these patterns can serve as references
to characterize structural heterogeneity of an unknown polymer translocating through a nanopore.
We hope the results reported in this letter
will be helpful in deciphering translocating characteristic of bio-polymers observed experimentally.

\par
This research has been partially supported by a UCF College of Science Seed grant. We thank Profs. Gary Slater and 
Hendrick de Hann for useful discussions.  


\begin{thebibliography}{99}
\bibitem{Unpublished}R. Adhikari and A. Bhattacharya, Unpublished.\\
In the pore friction dominated regime upon decreasing the solvent viscosity the system
is closer to its equilibrium state and the translocation dynamics
is determined by entropic barrier crossing.
This entropic barrier
is smaller for a stiffer chain and becomes negligible for rod-like polymer.
At low viscosity, the force due to pore friction increases as the solvent viscosity
decreases resulting  in an increase in the MFPT which explains the minimum of the MFPT at a particular viscosity $\gamma \sim \gamma_m$.
\bibitem{FENE1d}We have simulated driven translocation of a one dimensional compressible chain and find that the MFPT with the elastic constant $k_F$ increases as $\langle \tau \rangle \sim N^2k_F^{-1/8}$.
\end{thebibliography}
\end{document}